Superconductors that do not expel magnetic flux

Y. Tanaka[a*], H. Yamamori[a], T. Yanagisawa[a], T. Nishio[b], and S. Arisawa[c]

[a]National Institute of Advanced Industrial Science and Technology (AIST), Tsukuba, Ibaraki 305-8568, Japan
[b]Department of Physics, Tokyo University of Science, Shinjuku, Tokyo 162-8601, Japan
[c]National Institute for Materials Science, Tsukuba, Ibaraki 305-0047, Japan

*Corresponding author:   Telephone and Fax No.: +81-298-61-5720
                        E-mail address: y.tanaka@aist.go.jp

Abstract
An ultrathin superconducting bilayer creates a coreless fractional vortex when only the second layer has a hole. The quantization is broken by the hole, and the normal core disappears. The magnetic flux is no longer confined near the normal core, and its density profile around the hole becomes similar to that of a crème caramel; the divergence of the magnetic flux density is truncated around the center. We propose basic design of a practical device to realize a coreless fractional vortex.

1. Introduction

In a superconductor, a vortex concentrates its self-energy around its center [1]. For example, the region within a 2-μm-diameter circle holds more than 80% of self-energy, assuming a coherent length of 40 nm and a magnetic penetration depth of 80 nm for a 20-nm-thick niobium thin film [2-4]. This condition occurs because of quantization, in which the superconducting phase should rotate by $2\pi$ radians around the center, which provides a large amount of velocity to the superconducting pair near the center. If we can eliminate the singular point of the superconducting phase at the center (which corresponds to the normal core), we can expect a large reduction in the self-energy. This elimination can be made possible by introducing an inter-band phase difference soliton [5-32], which makes the phases of the first and second bands different and creates a fraction vortex [33-67]. The superconducting quantum phase does not rotate in one band while rotating by $2\pi$ radians in another band in the fractional vortex [30,33,43,65,66,68]. If we remove the second band around the center of the fractional vortex, the singular point is also removed. Elimination of the second band can be realized using a mimic two-band superconductor composed of an ultrathin superconducting bilayer [7, 69]. (The thickness of the layer is lesser than the penetration depth, and a zero-thickness limit yields a pure two-band superconductor with a substantially small inter-band Josephson interaction.) The first and second layers correspond to the first and second bands, respectively. The formation of a hole in the second layer (as shown in Fig. 1) corresponds to the elimination of the superconducting component in the second band. When a superconducting current streamline crosses the edge of the hole, as shown in Fig. 1(b), the quantization in the second layer is not constrained (because the path is not closed in the second layer). The first-layer phase still rotates by 0 radian along the streamline, which implies that if the center of the vortex is on the edge, as shown Fig. 1(b), the quantization constraint no longer forces the velocity of the superconducting pair to be large, and the velocity can thus be zero. To realize energy gain by the elimination of the core with the creation of a phase difference between the layers, some quantity must be traded off. One trade-off is the use of Josephson energy. Another one is the use of kinetic energy accompanied with an inter-layer circulating current that flows in the second layer, flows back to the first layer, and passes through the Josephson barrier [indicated by the yellow arrows in Fig. 1 (a)]. The magnetic induction of the inter-layer circulating current is mutually cancelled and can be ignored in the mimicked two-band superconductor limit. When these factors are satisfied because of the gain resulting from the elimination of the vortex core, a fractional vortex appears.

In this report, we present a basic design of a practical device to realize a coreless fractional vortex. In addition, we calculate the magnetic-flux-density profile, which reveals that the magnetic flux spreads around the hole and is no longer confined to a small area, as shown in Fig. 2.

## 2. Model and calculation method

We consider two specially overlapped thin layers, as shown in Fig. 1(a). These layers are coupled to each other through the Josephson interaction (where the critical Josephson current density is $j_c$), which is the limit of an ultrathin bilayer and is regarded as a two-band superconductor [7,69,70]. The current flows within a two-dimensional sheet. In each sheet, the current density does not vary along the direction perpendicular to the surface, as used in the approximation analyzing a Pearl vortex in a thin superconductor film [71-76]. However, the direction of the current in one layer can differ from that in the other layer [7, 69,70].

Brandt and Clem proposed in 2004 a calculation method for the current-density and magnetic-flux-density distributions of a thin circular disk using the London limit in which the superconducting pair density is constant [77] (Brandt presented another method in 2005 for arbitrary shapes with two-dimensional mesh [74]. However, for circular shapes, the Brandt-Clem method [77] with one dimensional mesh gives more accurate results with higher performance (reduction of computer resources). Thus, we adopted the Brand-Clem method.). First, to implement their method, we separate the current density into two components [68-70,78]: a circulating current density around the center of the vortex $(j_i^A)$, which induces a magnetic field and corresponds to the current density of the Brandt–Clem method, and an inter-layer circular current density $(j_i^B)$, which does not induce a magnetic field because of the mutual cancellation of the contributions from the two layers and can be ignored in the Brandt–Clem method. This approximation is practically the same as those adopted in the Brandt-Clem method, in which the sheet current density obtained by accumulating the local current density perpendicular to the film surface, determined the vector potential on the surface of the film in the zero-thickness limit [71,74,77]. The sheet current density of the circulating current around the center is assumed to be constant on the streamline according to the definition of the streamline of constant flow of incompressible fluid (In another view, it is Kirchhoff's first law.). Second, the current streamline is approximated by a circle, and the hole is shifted from the center of the disk to the perimeter of the hole to calculate the coreless vortex, as shown in Fig. 3, because the circulating current around the center is axially symmetric in the Brandt–Clem method. This is a variational function yielding required properties in the form of fractional and coreless vortex on the edge under our discussion. When this function provides lower energy than that of the conventional vortex at the normal location, and not on the edge, we can conclude with certainty that the coreless fractional vortex appears. An additive refinement, e.g., an ellipsoidal streamline, might reduce the energy further. However, the basic properties that are coreless and possess a fractional quantum, are expected to be retained.

By applying the London approximation to each layer indexed by $i$ [68-70,78], the gradient of the superconducting quantum phase ($\nabla\theta_i$) can be expressed using the velocity ($v_i$), mass ($m$), and charge ($e^*$) of a pair [68-70,78].

$$\hbar\nabla\theta_i = mv_i + e^*A \tag{1}$$

where $A$ is the vector potential and $\hbar = \frac{h}{2\pi}$, in which $h$ is the Plank constant. We define the two velocities for each layer as follows:

$$j_i^A \equiv ne^*v_i^A \tag{2a}$$
$$j_i^B \equiv ne^*v_i^B \tag{2b}$$

where $n$ is the pair density and $v_1^A = v_2^A$ corresponds to the velocity of the circular current in the two layers. Eq. (1) then becomes

$$\hbar\nabla\theta_i = mv_i^A + mv_i^B + e^*A. \tag{3}$$

The quantization condition for the first layer along the superconducting streamline yields [5,6,10,12,17,26,30,33,39-41,43,47,48,55,57,61,66,68,69]

$$0 = \oint \nabla\theta_1\, ds = \frac{1}{\hbar}\oint mv_1^A\, ds + \oint \frac{e^*}{\hbar}A\, ds + \frac{1}{\hbar}\oint mv_1^B\, ds. \tag{4}$$

The final term gives the phase shift due to the interlayer circulating current.

$$\hbar\Theta_s \equiv \oint mv_1^B\, ds \tag{5}$$

$v_1^B = 0$ under the hole. By introducing penetration depth $\lambda = \sqrt{\frac{1}{\mu_0}\frac{m}{n|e^*|}}$, where $\mu_0$ is the magnetic permeability of vacuum, Eq. (4) becomes

$$0 = \oint ne^*v_1^A\, ds + \frac{\oint A\, ds}{\mu_0\lambda^2} - \frac{\Phi_0}{\mu_0\lambda^2}\frac{\Theta_s}{2\pi}. \tag{6}$$

$\Phi_0 \equiv \frac{h}{|e^*|}$ is the magnetic-flux quantum. The total sheet current density that contributes to the magnetic induction is $J = ne^*(d_1v_1^A + d_2v_2^A) = ne^*dv_1^A$, where $d_i$ is the layer thickness. We set $d_1 = d_2$ in this study. Then $d = 2d_1$ when two layers are present, and $d = d_1$ when only one layer is present. $J$ is constant on the streamline, and $\oint nev_1^A\, ds = J\oint\frac{1}{d}ds$. Following the approximation of the streamline of $J$ by the circle, we introduce the effective thickness $d_{eff} = \frac{\oint \frac{1}{d} ds}{2\pi r}$, where $r$ is the radius of the circle of the streamline. Provided that the vector potential is parallel to the tangential direction of the streamline and its norm is constant, we obtain

$$\mu_0 \Lambda_{eff} J = -\left[ -\frac{\Theta_s}{2\pi} \frac{\Phi_0}{2\pi r} + A \right]. \tag{7}$$

where $\Lambda_{eff} = \frac{\lambda^2}{d_{eff}}$. This equation corresponds to Eq. (2) of the Brandt–Clem method [77]. The current coupling to the electromagnetic field ($J$) can be calculated when $\Theta_s$ is given.

The phase shift of the inter-layer phase difference between two edges ($P_S$ and $P_E$ in Fig. 3) is $2\Theta_s$. The inter-layer phase difference at the center is $\pi$ radians [as shown in Fig. 1(b)], increases to $2\pi$ radians at one edge (denoted by $P_E$ in Fig. 3), and shrinks down to 0 radian at another edge (denoted by $P_S$) with the increase in the radius ($r$ in Fig. 3) of the streamline [Fig. 1(b)]. The inter-layer phase difference ranges from 0 to $2\pi$ radians when the path does not cross the edge of the hole after $P_S$ and $P_E$ meet at $P_1$ (Fig. 3). When the length of the current streamline becomes longer than that of an inter-layer phase difference soliton [in our case, its length ($L_{soliton} = \frac{\pi}{2\lambda} \sqrt{\frac{d\Phi_0}{\pi\mu_0 j_c}}$) is approximated to be 200 μm], an inter-layer phase difference soliton is formed; this inter-layer phase difference only occurs in this soliton [5-32,68-70,77-79]. $L_{soliton}$ is obtained by dividing the slope of the inter-band phase difference at the center of the soliton by $2\pi$ radian [69]. The phase shift between two edges is defined as a function of the radius of the streamline. If we want to avoid a divergent singularity of the current at the center or assign a change in the current-flow direction near the center, the phase shift should be proportional to the square of the radius of the streamline near the center. Higher order functions can be mixed when the phase shift approaches $2\pi$ radians. However, although it could connect the phase slip more smoothly, the electromagnetic energy increases. Therefore, we set the square function of the radius of the streamline between 0 and $2\pi$ radians as a trial function. The phase variation in the streamline is estimated to be linear when the length of the streamline is less than that of the soliton. Beyond the length of the soliton, the phase variation is estimated to be composed of a linear part with the same slope as that at $\pi$ radians of the soliton function and a flat part with 0 or $2\pi$ radians [5,18,30]. This approximation means that a phase difference exists in the double layer for the whole path when the streamline length is shorter than $L_{soliton}$. When the streamline length is longer than $L_{soliton}$, an inter-layer phase difference is present in the limited fragment of the path, and the length of this fragment is $L_{soliton}$. With these considerations, we express the $r$ dependence of $\Theta_s$ as $\Theta_s = \pi \frac{r^2}{r_A^2}$ for $r < r_A$ and $\pi$ for $r > r_A$. $r_A$ is the hole diameter ($r_A = r_0$, see Fig. 3) for $r_0 < \frac{L_{soliton}}{2\pi}$. Otherwise, $r_A = \frac{L_{soliton}}{2\pi}$.

To implement the Brandt–Clem procedure, we introduce a non-equidistant grid. The streamline radius ($r_k$) is given as follows:

$$r_k = r_a + \frac{(r_b - r_a)f(u_k)}{f(1)}, \tag{8a}$$

$$f(u_k) = 30u_k^7 - 70(1+b)u_k^6 + 42(2b + (1+b)^2)u_k^5 - 105b(1+b)u_k^4 + 70b^2 u_k^3, \tag{8b}$$

$$u_k = \left(k - \frac{1}{2}\right)/52 \tag{8c}$$

where $r_a$ is zero for a coreless vortex and is the coherent length of a conventional vortex. $b$ is the ratio of the minimum radius ($r_0$) where the streamline in the second layer is close to the disk radius. Its value is 0.25 for a coreless vortex and 0.125 for a conventional vortex. $k$ ranges from 1 to 52. These equations yield a fine grid near the center and edge of the disk and near $r_0$. This grid was determined, as the separation of the grid becomes dense near $r_a$, $r_b$ and $r_0$. In other words, $f(u) \propto \frac{\partial}{\partial u}(u^2(u-1)^2(u-b)^2)$. This is an extension of the grid that Brandt and Clem adopted [77].

The kinetic energy of the pair is given by

$$E_k = \iiint \left(\frac{nd}{2}m(v_1^A + v_1^B)^2 + \frac{nd}{2}m(v_2^A + v_2^B)^2\right)d^3x. \tag{9}$$

Because the condition in which the interlayer circulating current is cancelled ($j_1^B + j_2^B = 0$) eliminates the cross terms of $v_i^A$ and $v_i^B$, the kinetic energy from $v_i^A$ and that from $v_i^B$ are separated as follows [41,43,68,78,79]:

$$E_k^A = \iiint \left(\frac{nd}{2}m((v_1^A)^2 + (v_2^A)^2)\right)d^3x, \tag{10a}$$

$$E_k^B = \iiint \left(\frac{nd}{2}m((v_1^B)^2 + (v_2^B)^2)\right)d^3x. \tag{10b}$$

The electromagnetic energy is given by

$$E_m = \frac{\mu_0}{2}\iiint H^2 d^3x = \frac{1}{2}\iint JA\, d^2x, \tag{11}$$

where $H$ is the magnetic field strength. $E_k^A + E_m$ corresponds to the usual self-energy of the disk with the vortex [77].

$$E_k^A + E_m = \frac{\Phi_0}{2}\int \frac{\Theta_s(r)}{2\pi}J(r)dr. \tag{12}$$

Under the applied field $(B_a)$, the electromagnetic energy contribution can be described following the Brandt–Clem method [74,77].

$$E_k^A + E_m = -\frac{1}{2}(M_1 + M_2)B_a + \frac{\Phi_0}{2}\int \frac{\Theta_s(r)}{2\pi}(J_1 + J_2)dr, \tag{13}$$

where $M_1$ is the moment induced by the vortex and $M_2$ is that induced by the applied field. $J_1$ is the sheet current density due to the vortex, and $J_2$ is that due to the applied field. For the interlayer circulating current, we derive

$$mv_1^B = -mv_2^B = \frac{\hbar\Theta_s}{l_s}. \tag{14}$$

$l_s$ is the length of the path where a phase difference exists between two layers in the double-layered region. Equations (10b) and (14) give the kinetic energy of the pair of the inter-layer circulating currents. To build up the inter-layer circulating current, the phase difference $(\varphi)$ ranges from $\pi - \Theta_s$ to $\pi + \Theta_s$. We approximate the phase as linearly varying along the streamline. The Josephson interaction can be given as

$$E_J = \frac{\Phi_0}{2\pi}j_c \iint dS(1 - \cos\varphi), \tag{15}$$

where $j_c$ is the Josephson critical current density. The total energy of the disk with the vortex can be estimated as follows:

$$E_{Disk} = E_k^A + E_m + E_k^B + E_J . \tag{16}$$

We estimate that $-\frac{1}{2}M_2B_a$ is the same for three cases—a disk with a conventional vortex, that with a coreless vortex, and that without any vortex. This estimation is based on the fact that the magnetic moment induced by the current in the film becomes the sum of the moment by the current accompanying the flux of the vortex and the moment by the current induced by the external field, as mentioned by Brandt [74,77]. We shift the center of the hole so that the center of the vortex is at the center of the disk when we estimate the energy, magnetic-flux-density distribution, and current-density distribution under the axially symmetric approximation.

## 3. Results

We briefly introduce the result for one example which is the most feasible one from the viewpoint of current device technology, which we can utilize as the first step to realize the coreless fractional vortex based on our model. It is one model case to demonstrate the effectiveness of our basic design. We can modify the design with varying penetration depth by changing the impurity doping to achieve more ideal thin-film limit. However, we observed that the current parameter is sufficient to realize the fractional vortex in preliminary experimental results [80].

Fig. 4 shows the energy difference between the disks with a vortex (conventional case), that with a fractional vortex, and that without a vortex where we approximate the diamagnetic magnetization induced by the external field to be equal. The diameters of the disk and hole are 80 and 10 μm, respectively. The thickness of one layer is 20 nm, and the Josephson current density is set to $10^5$ A/cm². The coherent length and penetration depth are set to 40 and 80 nm, respectively [3, 4]. The energy of the conventional vortex includes the energy of the normal core [69], which is approximately 8% of the total energy at 0 A/m. The self-energy of the coreless vortex is less than half of that of the conventional vortex at 0 A/m, which validates the discussions in the previous paragraphs. By increasing the field, the interaction between the magnetic moment of the fractional vortex and applied field decreases the energy, which becomes lower than that of the disk with or without a conventional vortex between 18 and 68 A/m.

The selected parameters are determined considering the practical situation. In principle, a large Josephson current and a large sample are unfavorable for creating a fractional vortex, because we should pay a large cost to create the $i$-soliton spanning a fractional vortex to the outer rim of the sample. The formation of the fractional vortex is more easily achieved with lower Josephson current density and a small sample. However, there is a limit coming from the current technology. To fabricate a reliable device, the Josephson current density should be large [81]. To obtain a reliable magnetic image by a scanning SQUID microscope, the size of the device should be certainly large also [82,83]. The selected conditions satisfy these requirements entirely.

## 4. Discussion

In this section, we discuss which experimental condition is appropriate to create a coreless fractional vortex in the model case. Then we make a remark on a noticeable knowledge in quantum science demonstrated by this experiment.

To realize a coreless vortex, the sample is cooled down under an applied field from a temperature higher than the transition temperature. At low temperatures, the coreless vortex is stable at the applied field and continues to exist after the external field is turned off [69,79].

Fig. 5 shows the magnetic-flux-density profile. A flat magnetic-flux-density profile can be realized when no constraint is present to force phase rotation. The current density reaches zero at the center. We note that the hole is still filled by the superconducting condensation of the first layer. In this model, the magnetic field also spreads outside the hole where two layers exist. The Higgs mechanism [84,85] still functions; however, the given 'Higgs mass' cannot confine the magnetic flux. (The superconducting condensation corresponds to the Higgs scalar field in superconductors.) In the absence of the quantization condition, the superconducting condensation cannot expel the magnetic field, although expulsion of the magnetic field is a fundamental superconducting property, known as the Meissner–Ochsenfeld effect [86]. London required quantization to explain the Meissner–Ochsenfeld effect [87]. When a sample does not have a hole, the London gauge ($j = -\frac{A}{\mu_0 \lambda^2}$) assures that a winding number of superconducting phase going around any closed path on the surface of the sample becomes zero. W. F. Edwards tried to explain the Meissner–Ochsenfeld effect without quantization in 1981 [88,89], but his conclusion was completely negated by J. B. Taylor and others in 1982 [90-94]. They discussed if Meissner–Ochsenfeld effect is really a quantum effect, by using a classical system [88-96]. Our study shows that the Meissner–Ochsenfeld effect is missed even in charged quantum condensations by damaging the boundary condition. For technical understanding, we may explain this phenomenon as follows. The current in the Lagrangian for Maxwell formalism cannot be replaced by the vector potential as seen in Eq. (7), though this replacement (with multiplying an appropriate constant) based on London consideration [87] usually reproduces Higgs' Lagrangian [84,85]. The reason why the vector potential cannot replace the current is that there is another current accompanying the soliton. The path for this soliton current is disconnected and not closed. Because of the phase shift due to the soliton current, $\frac{1}{2\pi}\int \nabla \theta_2 \, ds$ is not quantized to neither unit one nor a specific fractional value but varies continuously with increasing radius of streamline. It is noteworthy that two types of currents can be considered as reminiscent of the non-abelian current of two-band superconductors [47, 97].

    The coreless vortex can appear when we completely remove the superconductor passing through a hole. It does not conflict with the Higgs mechanism because there is no Higgs scalar field inside the hole and the connectivity of the superconducting quantum phase is not required inside the hole. The vortex has a unit flux quantum, because the phase must be connected outside the hole. When the superconductor remains as seen in a blind hole in a single layer, any coreless fractional vortex does not appear [98-105]. Other cases should also be mentioned. The amount of the magnetic flux can be smaller than a unit flux quantum for a mesoscopic sample with a finite circulating current at its outer rim [106]. It is

rather trivial that a ring has a zero or finite current when the amount of the magnetic flux inside the ring becomes a unit or fractional flux quantum. The Little-Parks experiment is based on this principle where the transition temperature or the current oscillates depending on the total magnetic flux inside the ring [12,19,107,108]. The period of the oscillation is quantized in this case. The "fractional vortex" in a mesoscopic ring is rather categorized into the extension of the Little-Parks experiment where a ring has a finite current. The fractional quantization can also be achieved by the externally applied current [109,110]. These currents directly couple to the electromagnetic field through the vector potential and these electromagnetically active currents rotate the phase. In our case, these currents are not required. Instead, a reminiscent of the non-abelian current (current accompanying the soliton and not directly coupling to the electromagnetic field) compensates the phase shift by the vector potential. The mechanism of fractionalization is rather similar to that discussed in spin-triplet $p$-wave quantum condensates [111, 112, 113], multi-component Bose-Einstein condensates [114,115] and particle physics [29,61,68,116]. There are theoretical reports on the fractional vortex for layered systems and multiband superconductors [10,12,16,24,31,33, 37, 38, 40, 42-44, 47, 48, 50-60, 65-69, 117-119], for example, inset 1 of Fig. 11 of ref. 53. Our report suggests that a partial suppression or elimination of one component enhanced these effects.

## 5. Conclusion

This letter may propose one method of studying quantum physics. Abrikosov predicted the quantized vortex concentrating the magnetic field around its center based on a phenomenological theory without experimental knowledge [1]. Our model sheds light on how to disperse the magnetic field concentrated by an Abrikosov quantized vortex. The ultrathin superconducting bilayer becomes a new platform to break the quantization condition without a quantum-to-classical crossover. The realized state resembles that of a perfect conductor with infinite conductivity but does not possess quantum properties [120]. Because the superconducting condensation is considered a Higgs scalar field [81], we speculate if a large object could possibly exist in a low-energy density in particle physics, such as dark energy [121], which corresponds to a coreless vortex with low magnetic-flux density.

Figures

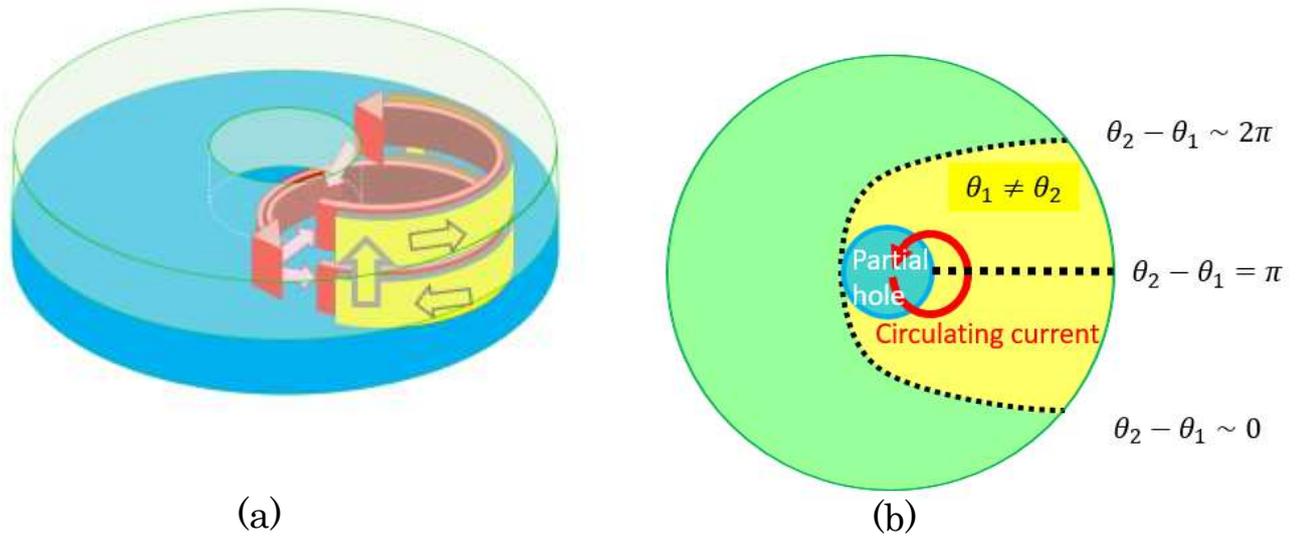

(a)  (b)

**Figure 1. Basic design of a device to promote a coreless fractional vortex.** (a) Perspective view. The structure is composed of ultrathin bilayers where one layer has a hole, but the other layer does not. These two layers are coupled to each other through an inter-layer Josephson interaction. The red arrows indicate a circulating current around the center of the coreless vortices. The yellow arrows indicate the inter-layer circulating current through the Josephson barrier. (b) Top view. $\theta_1$ and $\theta_2$ are the superconducting phases of the first and second layers, respectively.

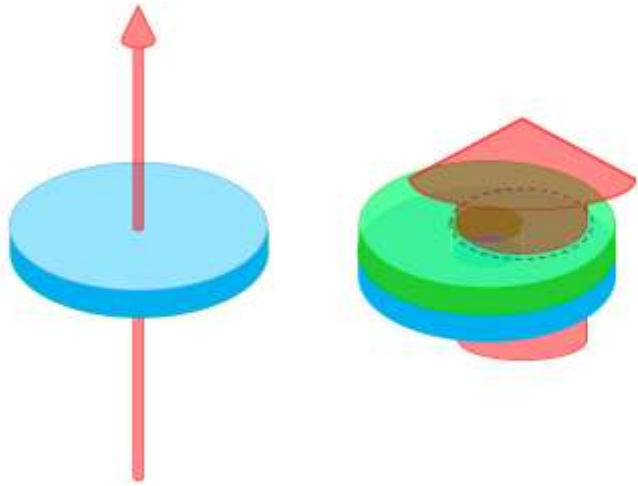

**Figure 2**. **Schematic illustration of the magnetic flux.** A sharply confined flux (denoted by red allows) for a single-layer becomes a broad flux (with low density) after attaching to another layer with a hole.

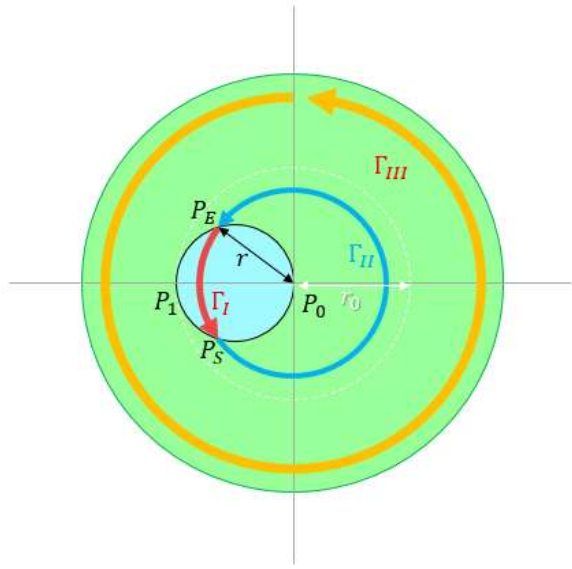

**Figure 3. Illustration of the approximation of the calculation.** The streamline is composed of $\Gamma_I$ and $\Gamma_{II}$ when it crosses the edges ($P_S$ and $P_E$) of the hole. When the radius of the circular current is larger than the diameter ($r_0$) of the hole, the current streamline does not cross the edges, as shown by $\Gamma_{III}$.

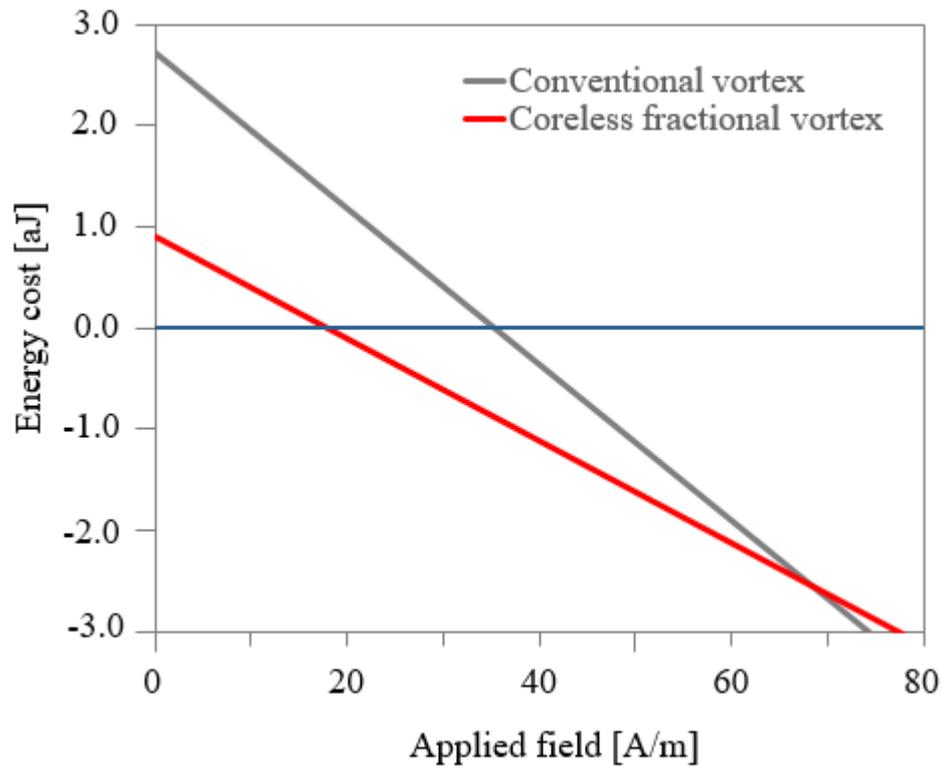

**Figure 4**. **Energy difference between the disks with and without a vortex.** The contribution of the interaction between the diamagnetic moment and applied field is approximated to be equal.

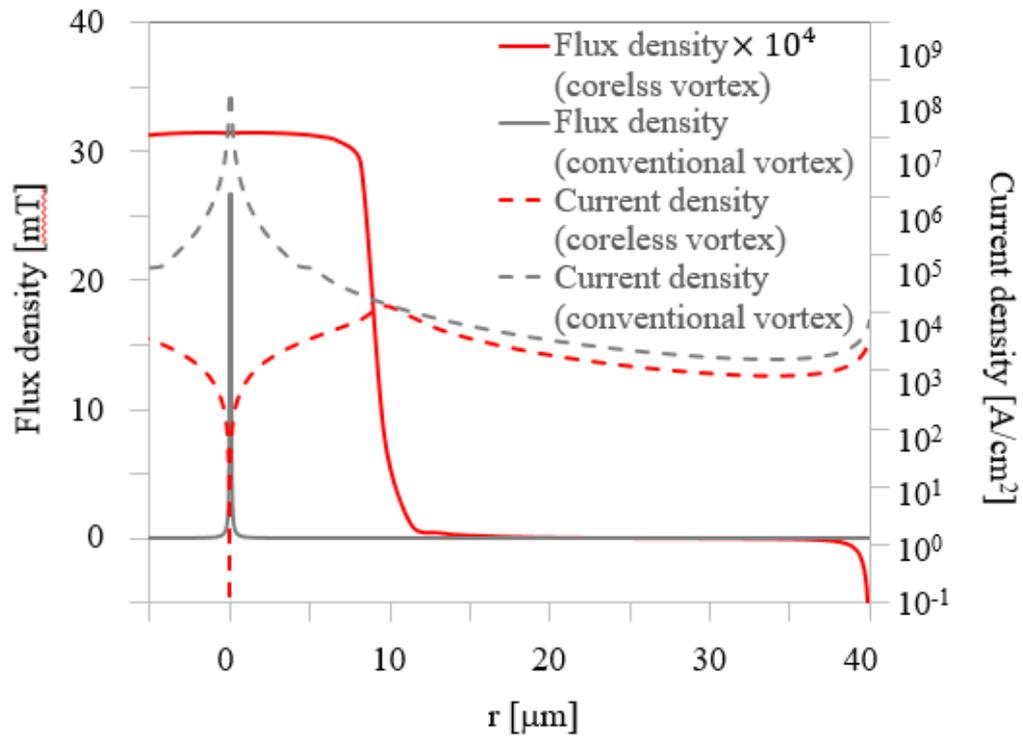

**Figure 5**. **Magnetic-flux-density and current-density distributions in the disk.** The current density is estimated at the location where two layers are present. The illustration of the flux density for the coreless vortex is magnified $10^4$ times.

**Acknowledgements:** We would like to thank Editage (www.editage.jp) for English language editing.

**Funding:** This work was partially supported by JSPS KAKENHI Grant Number JP 16K06275 and TIA collaborative research program 'Kakehashi'; Project Name: 'A feasibility study of fractional quantization'.

References

1. On the magnetic properties of superconductors of the second group. A. A. Abrikosov, Sov. Phys. JETP **5** (1957) 1174–1183.
2. Vortex–vortex interaction in thin superconducting films. E. H. Brandt. Phys. Rev. B **70** (2009) 134526.
3. Magnetic penetration depth measurements of superconducting thin films by a microstrip resonator technique. B. W. Langley, S. M. Anlage, R. F. W. Pease, M. R. Beasley. Rev. Sci. Instrum. **62** (1991) 1801–1812.
4. Superconducting properties of high-purity niobium. D. K. Finnemore, T. F. Stromberg, C. A. Swenson. Phys. Rev. **149** (1966) 231–243.
5. Phase instability in multi-band superconductors. Y. Tanaka. J. Phys. Soc. Jpn. **70** (2001) 2844–2847.
6. Soliton in two-band superconductor. Y. Tanaka. Phys. Rev. Lett. **88** (2002) 017002.
7. Interband phase modes and nonequilibrium soliton structures in two-gap superconductors. A. Gurevich, V. M. Vinokur, Phys. Rev. Lett. **90** (2003) 047004.
8. Anomalous effects of two gap superconductivity in $MgB_2$. A. Gurevich. Braz. J. Phys. **33** (2003) 700–704.
9. Dual superconductors and SU(2) Yang-Mills. J. High Energy Phys. 08 (2004) 035.
10. Vortices carrying an arbitrary fraction of magnetic flux quantum, neutral superfluidity and knotted solitons in two-gap Ginzburg–Landau model. E. Babaev. Physica C **404** (2004) 39-43
11. Flux flow and pinning of the vortex sheet structure in a two-component superconductor. Y. Matsunaga, M. Ichioka, K. Machida. Phys. Rev. B 60 (2004) 100502 (R).
12. Magnetic response of mesoscopic superconducting rings with two order parameters. H. Bluhm, N. C. Koshnick, M. E. Huber, K. A. Moler. Phys. Rev. Lett. **97** (2006) 237002. (Erratum: Phys. Rev. Lett. **98** (2007) 209902).
13. Phase textures induced by dc-current pair breaking in weakly coupled multilayer structures and two-gap superconductors. A. Gurevich, V. M. Vinokur. Phys. Rev. Lett. **97** (2006) 13700.
14. Ambiguity in the statistics of single-component winding vortex in a two-band superconductor. Y. Tanaka and A. Crisan, Physica B **404** (2008) 1033-1039.
15. $MgB_2$ thin films. X. X. Xi, Supercond. Sci. Technol. **22** (2009) 043001.
16. Soliton states in mesoscopic two-band-superconducting cylinders. S. V. Kuplevakhsky, A. N. Omelyanchouk, Y. S. Yerin. Low Temp. Phys. **37** (2011) 667-

677.

17. Chiral ground state in three-band superconductors. Y. Tanaka, T. Yanagisawa. J. Phys. Soc. Jpn. **79** (2010) 114706.

18. Ground state, collective mode, phase soliton and vortex in multiband superconductors. S.-Z. Lin. J. Phys.: Condens. Matter **26** (2014) 493202.

19. Observation of quantum oscillations in a narrow channel with a hole fabricated on a film of multiband superconductors. Y. Tanaka, G. Kato, T. Nishio, S. Arisawa. Solid State Commun. **201** (2015) 95–97.

20. Time-reversal symmetry-breaking in two-band superconductors. Y. Tanaka, P.M. Shirage, A. Iyo. Physica C: Superconductivity, **470** (2010) 2023–2026.

21. Topology of two-band superconductors. Y. Tanaka, A. Iyo, K. Tokiwa, T. Watanabe, A. Crisan, A. Sundaresan, N. Terada. Physica C **470** (2010) S966–S967.

22. Domain in multiband superconductors. Y. Tanaka, T. Yanagisawa, A. Crisan, P. M. Shirage, A. Iyo, K. Tokiwa, N. Nishio, A. Sundaresan, N. Terada. Physica C **471** (2011) 747–750.

23. Topological solitons in three-band superconductors with broken time reversal symmetry. J. Garaud, J. Carlström, E. Babaev. Phys. Rev. Lett. **107** (2011) 197001.

24. Topological defect-phase soliton and the pairing symmetry of a two-band superconductor: role of the proximity effect. V Vakaryuk, V Stanev, WC Lee, A Levchenko. Phys. Rev. Lett. **109** (2012) 227003.

25. Stability and Josephson effect of time-reversal-symmetry-broken multicomponent superconductivity induced by frustrated intercomponent coupling. X. Hu, Z. Wang. Phys. Rev. B **85** (2012) 064516.

26. Phase solitons in multi-band superconductors with and without time-reversal symmetry. S.-Z. Lin, X. Hu. New J. Phys. **14** (2012) 063021.

27. Phase solitons and subgap excitations in two-band superconductors. K. V. Samokhin. Phys. Rev. B **86** (2012) 064513.

28. Domain walls and their experimental signatures in s+is superconductors. J. Garaud, E. Babaev. Phys. Rev. Lett. **112** (2014) 017003.

29. Vortices and other topological solitons in dense quark matter. M. Eto, Y. Hirano, M. Nitta, S. Yasui. Prog. Theor. Exp. Phys. **2014** (2014) 012D01.

30. Multicomponent superconductivity based on multiband superconductors. Y. Tanaka. Supercond. Sci. Technol. **28** (2015) 034002.

31. Fractional flux plateau in magnetization curve of multicomponent superconductor loop. Z. Huang, X. Hu. Phys. Rev. B **92** (2015) 214516.

32. Soliton induced critical current oscillations in two-band superconducting bridges. P.

M. Marychev, D. Yu. Vodolazov. arXiv:1712.01529.

33. Vortices with fractional flux in two-gap superconductors and in extended Faddeev model. E. Babaev. Phys. Rev. Lett. **89**, (2002) 067001.

34. Phase diagram of planar U(1)×U(1) superconductor: Condensation of vortices with fractional flux and a superfluid state. E. Babaev. Nucl. Phys. B **686** (2004) 397–412.

35. Topological aspects of dual superconductors. S.-T. Hong, A. J. Niemi. Phys. Rev. B **70** (2004) 134511.

36. Field- and temperature-induced topological phase transitions in the three-dimensional N-component London superconductor. J. Smiseth, E. Smørgrav, E. Babaev, and A. Sudbø. Phys. Rev. B **71** (2005) 214509.

37. Dissociation of vortex stacks into fractional-flux vortices. A. De Col, V. B. Geshkenbein, G. Blatter. Phys. Rev. Lett. **94** (2005) 97001.

38. Domain walls and textured vortices in a two-component Ginzburg–Landau model. S. Madsen, Yu. B. Gaididei, P. L. Christiansen, N. F. Federsen. Phys. Lett. A. **344** (2005) 432-440.

39. Vortex state in a two-gap superconductor. J. Goryo, S. Soma, H. Matsukawa. Physica C **437–438** (2006) 86–88.

40. Non-Abrikosov vortex and topological knot in two-gap superconductors. Y. M. Cho, P. Zhang. Phys. Rev. B **73** (2006) 180506(R).

41. Vortex pinning in two-gap superconductors. J. Goryo. T. Saito, H. Matsukawa. J. Phys.: Conf. Ser. **89** (2007) 012022.

42. Thermodynamically stable noncomposite vortices in mesoscopic two-gap superconductors. L. F. Chibotaru, V. H. Dao, A. Ceulemans. Euro. Phys. Lett. **78** (2007) 47001.

43. Deconfinement of vortices with continuously variable fractions of the unit flux quanta in two-gap superconductors. J. Goryo, S. Soma, H. Matsukawa. EPL **80** (2007) 17002.

44. Anomalous AC susceptibility response of $(Cu,C)Ba_2Ca_2Cu_3O_y$: Experimental indication of two-component vortex matter in multi-layered cuprate superconductors. A. Crisan, Y. Tanaka, D. D. Shivagan, A. Iyo, L. Cosereanu, K. Tokiwa, T. Watanabe. Jpn. J. Appl. Phys. **46** (2007) L451–L453.

45. Two-band superconductor magnesium diboride, X. X. Xi, Rep. Prog. Phys. **71** (2008) 116501.

46. Two-dimensional vortex behavior in highly underdoped $YBa_2Cu_3O_{6+x}$ observed by scanning Hall probe microscopy. J. W. Guikema, H. Bluhm, D. A. Bonn, R. Liang, W. N. Hardy, K. A. Moler. Phys. Rev. B **77** (2008) 104515.

47. Topological objects in two-gap superconductor. Y. M. Cho and P. M. Zhang. Euro. Phys. J. B **65** (2008) 155-178.
48. Vortices in two-gap superconductors. J. Goryo, T. Saito, H. Matsukawa. J. Phys. Soc. Jpn. **77** (2008) Suppl. A, 272–274.
49. Magnetic force microscopy study of interlayer kinks in individual vortices in the underdoped cuprate superconductor $YBa_2Cu_3O_{6+x}$. L. Luan, O. M. Auslaender, D. A. Bonn, R. Liang, W. N. Hardy, K. A. Moler. Phys. Rev. B **79** (2009) 214530.
50. Vortex molecule and $i$-soliton studies in multilayer cuprate superconductors. D. D. Shivagan, A. Crisan, P. M. Shirage, A. Sundaresan, Y. Tanaka, A. Iyo, K. Tokiwa, T. Watanabe, N. Terada. J. Phys.: Conf. Ser. **97** (2009) 012212.
51. Magnetic field delocalization and flux inversion in fractional vortices in two-component superconductors. E. Babaev, J. Jäykkä, M. Speight Phys. Rev. Lett. **103** (2009) 237002.
52. Stable fractional flux vortices in mesoscopic superconductors. L. F. Chibotaru, V. H. Dao. Phys. Rev. B **81** (2010) 020502(R).
53. Vortex matter in mesoscopic two-gap superconducting disks: Influence of Josephson and magnetic coupling. R. Geurts, M. V. Milošević, F. M. Peeters. Phys. Rev. B **81** (2010) 214514.
54. Vortex matter in mesoscopic two-gap superconductor square. P. J. Pereira, L. F. Chibotaru, and V. V. Moshchalkov, Phys. Rev. B **84** (2011) 144504.
55. Exotic vortex matter: pancake vortex molecules and fractional-flux molecules in some exotic and/or two-component superconductors, A. Crisan, Y. Tanaka, A. Iyo. J. Supercond. Nov. Magn. **24**(2011)1-6.
56. Stability of fractional vortex states in a two-band mesoscopic superconductor. J. Piña, C. C. de Souza Silva, M. V. Milošević. Phys. Rev. B **86** (2012) 024512.
57. Vortices and chirality in multi-band superconductors. T. Yanagisawa, Y. Tanaka, I. Hase, K. Yamaji. J. Phys. Soc. Jpn. **81** (2012) 024712.
58. Half-flux quanta and vortex-antivortex pairs in two-band superconductors. L. Wen, G.-Q. Zha, S.-P. Zhou. Euro. Phys. Lett. **102** (2013) 27004.
59. Dissociation transition of a composite lattice of magnetic vortices in the flux-flow regime of two-band Superconductors. S.-Z. Lin and L. Bulaevskii. Phys. Rev. Lett. **110** (2013) 087003.
60. Stabilizing fractional vortices in multiband superconductors with periodic pinning arrays. S.-Z. Lin, C. Reichhardt. Phys. Rev. B **87** (2013) 100508(R).
61. Quarks and fractionally quantized vortices in superconductors: an analogy between two worlds. T. Yanagisawa. in Recent Advances in Quarks Research. ed. H Fujikage and K Hyobanshi (Nova Science, New York, 2013) 113–146.
62. Optimizing mesoscopic two-band superconductors for observation of fractional vortex states. J. C. Piña, C. C. de Souza Silva, M. V. Milošević´. Physica C **503** (2014) 48–51.
63. Fractional vortex molecules and vortex polygons in a baby Skyrme model. M. Kobayashi, M. Nitta. Phys. Rev. D 87 (2013) 125013.
64. Vortex states in mesoscopic three-band superconductors. S. Gillis, J. Jäykkä, M. V. Milošević. Phys. Rev. B **89** (2014) 024512.
65. Distinct magnetic signatures of fractional vortex configurations in multiband superconductors. R. M. da Silva, M. V. Milošević, D. Dominguez, F. M. Peeters, J. A. Aguiar. Appl. Phys. Lett. **105** (2015) 232601.
66. Dipole magnetic flux in two-gap superconducting thin film in Ginzburg–Landau theory. J. Zhou, Z. Gan. Physica C **510** (2015) 42-47.
67. Nonequilibrium interband phase textures induced by vortex splitting in two-band superconductors. A. S. Mosquera, Polo, R. M. da Silva, A. Vagov, A. A. Shanenko, C. E. Deluque Toro, J. Albino Aguiar. Phys. Rev B **96** (2017) 054517.
68. Interpretation of abnormal AC loss peak based on vortex-molecule model for a multicomponent cuprate superconductor. Y. Tanaka, A. Crisan, D. D. Shivagan, A. Iyo, K. Tokiwa, T. Watanabe. Jpn. J. Appl. Phys. **46** (2007) 134–145.
69. Decomposition of a unit quantum and isolation of a fractional quantum by an externally injected soliton in an ultrathin superconducting bi-layer. Y. Tanaka, H. Yamamori, T. Yanagisawa, T. Nishio, S. Arisawa. Physica C **538** (2017) 12–19.
70. Voltage-less alternating current (AC) Josephson effect in two-band superconductors. Y. Tanaka, H. Yamamori, T. Yanagisawa, T. Nishio, S. Arisawa. Physica C **538** (2017) 6-11.
71. Current distribution in superconducting films carrying quantized fluxoids. J. Pearl. Appl. Phys. Lett. **5** (1964) 65–66.
72. Two-dimensional vortices in a stack of thin superconducting films: A model for high-temperature superconducting multilayers. J. R. Clem. Phys. Rev. B **43** (1991) 7837–7846.
73. Josephson junction in a thin film. V. G. Kogan, V. V. Dobrovitski, J. R. Clem, Y. Mawatari, R. G. Mints. Phys. Rev. B **63** (2001) 144501.
74. Thin superconductors and SQUIDs in perpendicular magnetic field. E. H. Brandt. Phys. Rev. B **72** (2005) 024529.
75. Observation of an extended magnetic field penetration in amorphous superconducting MoGe films. T. Nishio, S. Okayasu, J.-I. Suzuki, N. Kokubo, K.

Kadowaki. Phys. Rev. B **77** (2008) 052503.

76. Fundamental studies of superconductors using scanning magnetic imaging. J. R. Kirtley. Rep. Prog. Phys. **73** (2010) 126501.
77. Superconducting thin rings with finite penetration depth. E. H. Brandt, J. R. Clem. Phys. Rev. B **69** (2004) 184509.
78. Emergence of an interband phase difference and its consequences in multiband superconductors. Y. Tanaka. Springer Series in Materials Science **261** (2017) 185–218.
79. Current-induced massless mode of the interband phase difference in two-band superconductors. Y. Tanaka, I. Hase, T. Yanagisawa, G. Kato, T. Nishio, S. Arisawa. Physica C **516** (2015) 10-16.
80. Experimental formation of a fractional vortex in a superconducting bi-layer. Y. Tanaka, H. Yamamori, T. Yanagisawa, T. Nishio, S. Arisawa. Submitted to Physica C during communication of this article.
81. Current status and future prospect of the Nb-based fabrication process for single flux quantum circuits. M. Hidaka, S. Nagasawa, T. Satoh, K. Hinode, Y. Kitagawa. Supercond. Sci. Technol. **19** (2006) S138–S142.
82. Direct observation of local shielding currents in superconducting thin films under low magnetic field by scanning superconducting quantum interference device microscopy. S. Arisawa, K. Mochiduki, K. Yun, T. Hatano, I. Iguchi, K. Endo. Jpn. J. Appl. Phys. **51** (2012) 095804.
83. Observation of an extended magnetic field penetration in amorphous superconducting MoGe films. T. Nishio, S. Okayasu, J.-I. Suzuki, N. Kokubo, K. Kadowaki. Phys. Rev. B **77** (2008) 052503.
84. Broken symmetries and the masses of gauge bosons. P. W. Higgs. Phys. Rev. Lett. **13** (1964) 508–509.
85. Vortex-line models for dual strings. H. B. Nielsen, P. Olesen. Nucl. Phys. B **61** (1973) 45–61.
86. Ein neuer Effekt bei Eintritt der Supraleitfähigkeit. W. Meissner, R. Ochsenfeld. Naturwissenschaften **21** (1933) 787–788.
87. The electromagnetic equations of the superconductor. F. London, H. London. Proc. Roy. Soc. (London) **A149** (1935) 71–88.
88. Classical derivation of the London equations. Phys. Rev. Lett. **47** (1981) W. F. Edwards. 1863–1866.
89. Edwards Responds. W. F. Edwards. Phys. Rev. Lett. **49** (1982) 419–420.

90. Comments on the variational principle and superfluid mechanics discussion. S. Putterman. Phys. Lett. A **89** (1982) 146–148.
91. Distinction between a perfect conductor and a superconductor. F. S. Henyey. Phys. Rev. Lett. **49** (1982) 416.
92. Comment on "Classical derivation of the London equations". B. Segall, L. L. Foldy, R. W. Brown. Phys. Rev. Lett. **49** (1982) 417.
93. Comment on the applicability of Lagrangian methods to the London equations. Paul G. N. deVegvar. Phys. Rev. Lett. **49** (1982) 418.
94. A classical derivation of the Meissner effect? J. B. Taylor. Nature **299** (1982) 681–682.
95. Classical perfect diamagnetism: Explusion of current from the plasma interior. S. M. Mahajan. Phys. Rev. Lett. 100 (2008) 075001.
96. Relaxation of toroidal plasma and generation of reverse magnetic fields. J. B. Taylor. Phys. Rev. Lett. **33** (1974) 1139-1141.
97. Conservation of isotopic spin and isotopic gauge invariance. C. N. Yang, R. L. Mills. Phys. Rev. **96** (1954) 191-195.
98. Nucleation of vortices inside open and blind microholes. A. Bezryadin, Yu. N. Ovchinnikov, B. Pannetier. Phys. Reb. B 53 (1996) 8553-8560.
99. Superconducting vortex state in a mesoscopic disk containing a blind hole. G. R. Berdiyorov, M. V. Milošević, B. J. Baelus, F. M. Peeters. Phys. Rev. B **70** (2004) 024508.
100. Flux-pinning properties of superconducting films with arrays of blinds holes. S. Raedts, A. V. Sihanek, M. J. Van Bael, V. V. Moshchalkov. Phys. Rev. B **70** (2004) 024509.
101. Flux-pinning properties of holes and blind holes arranged periodically in a superconductor. S. Raedts, A. V. Sihanek, M. J. Van Bael, R. Jonckheere, V. V. Moshchalkov. Physica C **404** (2004) 298-301.
102. The structure and manipulation of vortex states in a superconducting square with 2 x 2 blind holes. G. R. Berdiyorov, M. V. Milošević, F. M. Peeters. J. Low Temp. Phys. **139** (2005) 229-238.
103. Anisotropic avalanches and flux penetration in patterned superconductors. D. G. Gheorghe, M. Menghini, R. J. Wijngaarden, S. Raedts, A. V. Silhanek, V. V. Moshchalkov. Physica C **437-438** (2006) 69-72.
104. Vortex structure and critical parameters in superconducting thinfilms with arrays of pinning centers. G. Berdiyorov. Doctoral Dissertation (University of Antwerp 2007).

105. Composite vortex ordering in superconducting films with arrays of blind holes. G. R. Berdiyorov, M. V. Milošević, F. M. Peeters. New J. Phys. **11** (2009) 013025.

106. Non-quantized penetration of magnetic field in the vortex state of superconductors. A. K. Geim, S. V. Dubonos, I. V. Grigorieva, K. S. Novoselov, F. M. Peeters. Nature **407** (2000) 55-57.

107. Observation of quantum periodicity in the transition temperature of a superconducting cylinder. W. A. Little, R. D. Parks. Phys. Rev. Lett.9 (1962) 9-12.

108. Fluctuation superconductivity in mesoscopic aluminum rings. N. C. Koshnick, H. Bluhm, M. E. Huber, K. A. Moler. Science 318 (2007) 1440-1443.

109. Superconducting multilayer technology for Josephson devices. J. M. Meckbach. Doctoral Dissertation (Karlsruher Institute für Technologie. 1994, ISBN 978-3-7315-0122-0).

110. Dynamics of semifluxons in Nb long Josephson 0-π junction. E. Goldobin, A. Sterck, T. Gaber, D. Koelle, R. Kleiner. Phys. Rev. Lett. **92** (2004) 057005

111. Vortices in rotating superfluid $^3$He. O. V. Lounasmaa, E. Thuneberg. PNAS 96 (1999) 7760–7767.

112. Observation of half-quantum vortices in topological superfluid $^3$He. S. Autti, V. V. Dmitriev, J. T. Mäkinen, A. A. Soldatov, G. E. Volovik, A. N. Yudin, V. V. Zavjalov, V. B. Eltsov. Phys. Rev. Lett. **117** (2016) 255301.

113. Vortex structure in superconductors with a many-component order parameter. Y. A. Izyumov, V. M. Laptev. Phase Transition 20 (1990) 95-112.

114. Vortices in a Bose-Einstein condensate. M. R. Matthews, B. P. Anderson, P. C. Haljan, D. S. Hall, C. E. Wieman, E. A. Cornell. Phys. Rev. Lett. **83** (1999) 2498-2501.

115. Domain walls of relative phase in two-component Bose-Einstein condensates. D. T. Son, M. A. Stephanov. Phys. Rev. A 65 (2002) 063621.

116. Center vortices, nexuses, and fractional topological charge. J. M. Cornwall. Phys. Rev. D **61** (2000) 085012.

117. Vortices and dissipation in a bilayer thin film superconductor. W. Zhang and H. A. Fertig, Phys. Rev. B **71** (2005) 224514.

118. Phase diagram of a lattice of pancake vortex molecules. Y. Tanaka, A. Crisan, D. D. Shivagen, A. Iyo, P. M. Shirage, K. Tokiwa, T. Watanabe, N. Terada. Physica C **469** (2009) 1129–1131.

119. Enhanced stability of vortex-antivortex states in two-component mesoscopic superconductors. Phys. Rev. B 87 (2013) 024501.

120. M. Tinkham, Introduction to Superconductivity, McGraw–Hill, Inc. New York,

1996, pp. 5 and 18.

121.   Prospects for probing the dark energy via supernova distance measurement. D. Huterer, M. S. Turner. Phys. Rev. D **60** (1999) 081301.